\documentclass[usenatbib]{mn2e}

\usepackage{amsmath}
\usepackage{graphicx}
\usepackage{txfonts}

\def\apj{ApJ}
\def\apjl{ApJL}

\def\aap{A\&A} 
\def\apss{Ap\&SS}

\def\mnras{MNRAS}

\def\ssr{Space~Sci.~Rev.}
\def\icarus{Icarus}
\def\aaps{A\&AS}
\def\apjs{ApJS}
\def\baas{BAAS} 
\def\planss{Planet.~Space~Sci.}

\bibpunct{(}{)}{;}{a}{}{,}

\title[Thermal reactivity of HCN and NH$_3$]{The thermal reactivity of HCN and NH$_3$ in interstellar ice analogues.}

\author[J.~A. Noble et al.]{\parbox{\textwidth}{J.~A. Noble,\thanks{Email: jennifer.noble@univ-amu.fr} 
P. Theule, 
F. Borget, 
G. Danger, 
M. Chomat, 
F. Duvernay, 
F. Mispelaer, and 
T. Chiavassa}\vspace{0.4cm}\\
\parbox{\textwidth}{Aix-Marseille Universit\'{e}, PIIM UMR-CNRS 7345, 13397, Marseille, France.}}
\begin{document}

\date{Accepted 1988 December 15. Received 1988 December 14; in original form 1988 October 11}
\pagerange{\pageref{firstpage}--\pageref{lastpage}} \pubyear{2002}

\maketitle

\label{firstpage}

\begin{abstract}  %200 words limit
HCN is a molecule central to interstellar chemistry, since it is the simplest molecule containing a carbon-nitrogen bond and its solid state chemistry is rich.
The aim of this work was to study the NH$_3$~+~HCN~$\rightarrow$~NH$_4^+$CN$^-$ thermal reaction in interstellar ice analogues.
Laboratory experiments based on Fourier transform infrared spectroscopy and mass spectrometry were performed to characterise the NH$_4^+$CN$^-$  reaction product and its formation kinetics.
This reaction is purely thermal and can occur at low temperatures in interstellar ices without requiring non-thermal processing by photons, electrons or cosmic rays. 
The reaction rate constant has a temperature dependence of $k(T)$ = 0.016$^{+0.010}_{-0.006}$~s$^{-1}\exp(\frac{-2.7\pm0.4\text{~kJ\,mol}^{-1}}{RT})$ when NH$_3$ is much more abundant than HCN. When both reactants are diluted in water ice, the reaction is slowed down. We have estimated the CN$^-$ ion band strength to be A$_{\text{CN}^-}$~=~$1.8\pm1.5\times10^{-17}$~cm\,molec$^{-1}$ at both 20~K and 140~K. NH$_4^+$CN$^-$ exhibits zeroth-order multilayer desorption kinetics with a rate of $k_{\text{des}}(T)$ = 10$^{28}$~molecules\,cm$^{-2}$\,s$^{-1}\exp(\frac{-38.0\pm1.4\text{~kJ\,mol}^{-1}}{RT})$.
The NH$_3$~+~HCN~$\rightarrow$~NH$_4^+$CN$^-$ thermal reaction is of primary importance because \textit{(i)} it decreases the amount of HCN available to be hydrogenated into CH$_2$NH, \textit{(ii)} the NH$_4^+$ and CN$^-$ ions react with species such as H$_2$CO, or CH$_2$NH to form complex molecules, and \textit{(iii)} NH$_4^+$CN$^-$ is a reservoir of NH$_3$ and HCN, which can be made available to a high temperature chemistry.
\end{abstract}

\begin{keywords}
astrochemistry -- ISM: molecules -- molecular processes -- molecular data -- methods: laboratory.
\end{keywords}

\section{Introduction}

The HCN molecule plays a fundamental role in interstellar chemistry as it is the simplest molecular species containing both C and N atoms (while the radical CN is the simplest interstellar species composed of C and N); reactions involving HCN can produce key species intermediate to the formation of amino acids, such as aminoacetonitrile \citep{Danger11}.

HCN has been observed in the gas phase in various environments including molecular clouds \citep{Snyder71,Clark74}, circumstellar envelopes \citep{Bieging84}, the comae of comets \citep{Ip90}, the atmospheres of Neptune \citep{Rosenqvist92} and Titan \citep{Tokunaga81}, and was recently discovered in the Murchison meteorite \citep{Pizzarello12}. Typical abundances with respect to H$_2$ are 10$^{-9}$~--~10$^{-8}$ towards molecular clouds \citep{Hirota98} and massive protostars \citep{Schreyer97}, rising to 10$^{-6}$ in the hot envelope preceding a hot core \citep{Lahuis00,Boonman01} due to a combination of gas phase chemistry and the sublimation of ices from dust grain surfaces. As such, HCN has been shown to be a linear tracer of ongoing star formation and is used to probe dense gas in star forming regions \citep{Hirota98}. The gas phase ratios between isomers such as HNC and HCN indicate the extent of photochemical isomerisation that has occurred in astrophysical environments. The HNC/HCN ratio is highly dependent on temperature, and has been observed to be $\sim$~1 in prestellar cores \citep{Padovani11,Herbst00}, 0.6~--~0.9 in infrared dark clouds \citep{Vasyunina11}, and less than 0.3 in Neptune's atmosphere \citep{Moreno11}. The gas phase ratio CN/HCN can change by orders of magnitude between the edge of a molecular cloud and the densest regions of the core, as it is sensitive to photodissociation channels \citep{Boger05}. A greater knowledge of the grain surface chemistry network between CN and HCN is needed to understand CN/HCN ratios in dense cores \citep{Chapillon11}. CN$^-$ is the first diatomic anion to have been positively identified in an astrophysical environment, having been observed recently in the gas phase in the C-star envelope IRC~+10216 \citep{Agundez10}.

One environment where HCN is of particular interest is in comets and their comae. The CN moiety (part) is key in the production of amino acids \citep{Ferris84}, recently observed in comets for the first time by the \emph{Stardust} spacecraft \citep{Elsila09}.  Both HCN and CN have been observed extensively in the gas phase on lines of sight towards cometary comae \citep[e.g.][]{Heubner74,Bockelee87,Woodney02}. HCN is believed to be the major source of the CN radical \citep{Paganini10}, although other sources, such as refractory organic compounds, are necessary in order to produce the observed CN abundances \citep{Fray05}. In the solid phase, HCN is unlikely to exist solely in its monomeric form, but also in the form of polymers and complex species which act as HCN reservoirs, such as hexamethyltetramine \citep[HMT, ][]{Bernstein95,Vinogradoff11} or poly(methylene-imine) \citep[PMI, ][]{Danger11}. Solid phase HCN has yet to be positively identified in comets, but a tentative detection of solid phase HCN on Triton was made recently using the AKARI telescope \citep{Burgdorf10}.

Experimentally, solid phase HCN has been produced under astrophysically relevant conditions upon irradiation of CH$_4$:N$_2$ mixtures with 5~keV electrons \citep{Jamieson09}. 
HCN can react as an electrophile, as in the reaction of HCN with H atoms in the solid phase which produces methylamine \citep{Theule11a}, or as a nucleophile, as in its reaction with H$_2$CO to form hydroxyacetonitrile HOCH$_2$CN \citep{Danger12}.
The participation of HCN in a low temperature solid phase Strecker synthesis (a series of reactions to synthesise an amino acid) yields aminoacetonitrile, NH$_2$CH$_2$CN, via thermal processing of an ice mixture of CH$_2$NH, NH$_3$ and HCN \citep{Danger11}. In addition, the irradiation of HCN in ice mixtures and aqueous solutions with photons or charged particles has been seen to yield complex ions \citep{Gerakines04}, amino acids \citep{Bernstein02}, amides, carboxylic acids and bases \citep{Colin09}. 

HCN is also a weak acid, therefore it could react with bases to produce salts. Previously, acid-base reactions have been shown to yield salts under astrophysically relevant conditions. The reaction HNCO + NH$_3$ produces NH$_4^+$OCN$^-$ \citep{Demyk98,Raunier03a,vanBroekhuizenAA04} with an activation energy of 0.4~kJ\,mol$^{-1}$ \citep{Mispelaer12}; HNCO + H$_2$O produces H$_3$O$^+$OCN$^-$  \citep{Raunier03b} with an activation energy of 26~kJ\,mol$^{-1}$ \citep{Theule11a}, which then further reacts to form the isomerisation product HOCN with an activation energy of 36~kJ\,mol$^{-1}$ \citep{Theule11b}; HCOOH + NH$_3$ yields NH$_4^+$HCOO$^-$ \citep{Schutte99}. Such species have relatively high sublimation temperatures (greater than $\sim$~150~K) compared with the simple molecules such as CO$_2$, H$_2$O and CO which compose the bulk of the icy mantles on dust grains \citep{Raunier04,Noble12}. 
As such, it is likely that salts will remain on the grain surface following heating processes that sublime the majority of the ice mantle, ultimately forming part of the refractory organic residue observed on comet grains \citep{GlavinMPS08}.

In this paper, we report the results of a study of the purely thermal reaction of the weak acid HCN with the base NH$_3$:

\begin{equation}\label{eqn_hcn_nh3}
 \mbox{HCN} + \mbox{NH}_3 \rightarrow \mbox{NH}_4^+\mbox{CN}^-.
\end{equation}

NH$_3$ has been observed in astrophysical ices at abundances of 2~--~15~\% H$_2$O \citep{Lacy98,Bottinelli10}, 
and is therefore a key basic, nucleophilic species available to react with acidic, electrophilic molecules in grain ice mantles. 

The 6.0 and 6.85~$\mu$m bands are observed in ices towards many astrophysical environments, including young stellar objects \citep[YSOs,][]{Keane01,Boogert08} and quiescent molecular clouds \citep{Boogert11}, but have yet to be fully characterised. 
NH$_4^+$ is considered to be a likely candidate to account for some of the feature at 6.85~$\mu$m \citep{Schutte03}. 
It has been suggested that most of the 6.0~$\mu$m excess, and at least some of the 6.85~$\mu$m band, is related to highly processed ices \citep{Gibb02}. However, given that the bands are observed in quiescent regions, some low-energy formation route to their carriers must be possible. Laboratory studies have shown that NH$_4^+$ can be produced via acid-base reactions in ices at 10~K with no irradiation \citep[e.g.][]{Raunier03a}.

Thus, reaction (\ref{eqn_hcn_nh3}) is important as it involves the abundant ice species NH$_3$, produces NH$_4^+$, and also involves HCN, the simplest molecule containing a CN bond, which could be a prerequisite for the formation of amino acids \citep{Danger11}.

In this work, the NH$_4^+$CN$^-$ product of the NH$_3$~+~HCN reaction is spectroscopically characterised using Fourier transform infrared (FTIR) spectroscopy. The kinetics of the NH$_3$~+~HCN reaction are investigated using isothermal experiments employing FTIR spectroscopy. The temperature dependence of the reaction rate is determined to be k(T) = 0.016$^{+0.010}_{-0.006}$~s$^{-1}$\,$\exp(\frac{-2.7\pm0.4\text{~kJ\,mol}^{-1}}{RT})$. The band strength of CN$^-$ is determined to be A$_{\text{CN}^-}$~=~1.8~$\pm$~1.5~$\times$~10$^{-17}$~cm\,molec$^{-1}$. Temperature programmed desorption (TPD) experiments using mass spectrometry allows calculation of the desorption rate as $k_{\text{des}}(T)$ =  10$^{28}$~molecules\,cm$^{-2}$\,s$^{-1}$\,$\exp(\frac{-38.0\pm1.4\text{~kJ\,mol}^{-1}}{RT})$ for a zeroth-order multilayer desorption. 
NH$_4^+$CN$^-$ is therefore more refractory than H$_2$O, and thus can participate in a warm, water-free chemistry.

\section{Experimental}\label{sec-expt}

Experiments were performed using the RING experimental set-up, as described elsewhere \citep{Theule11b}. 
Briefly, RING consists of a gold-plated copper surface within a high vacuum chamber (a few 10$^{-9}$~mbar). The temperature of the surface is controlled using a Lakeshore Model 336 temperature controller, a closed-cycle helium cryostat (ARS Cryo, model DE-204 SB, 4~K cryogenerator), and a heating resistor; the temperature is measured using a DTGS~670 Silicon diode, with a 0.1~K uncertainty.

\begin{table}
\setlength{\tabcolsep}{2pt}
\caption{Experiments performed in the current work. $^a$}
\label{table1}
\begin{center}
\begin{tabular}{p{1cm}lccccccccc}
\hline
ID     & Molecules                                      & \multicolumn{9}{l}{Ratio at deposition}                   \\
        &                                                       & HCN & : & NH$_3$ & : & H$_2$O & : & CO$_2$ & : & CO \\
\hline
i       & pure HCN                                       & 1     & :  &  0          & : & 0           & : & 0           & : & 0        \\% ;;EXPT 17 --> i
ii      & pure NH$_3$                                  & 0     & :  &  1          & : & 0           & : & 0           & : & 0       \\% ;;EXPT 18 --> ii
iii     & HCN \& NH$_3$                             & 1     & :  & 20        & : & 0           & : & 0           & : & 0       \\% ;; EXPT 4 --> iii
iv     & HCN \& NH$_3$                             & 1.5  & :  & 1           & : & 0           & : & 0           & : & 0         \\% ;;Expt 3 --> iv
v      & HCN \& H$_2$O                             & 1     & :  & 0           & : & 18         & : & 0           & : & 0          \\% ;;EXPT 0 --> v
vi     & HCN, NH$_3$ \& H$_2$O               &  1    & :  & 3           & : & 30         & : & 0           & : & 0            \\% ;;EXPT 8 --> vi
vii    & HCN, NH$_3$, H$_2$O \& CO$_2$ & 2.5  & : &  6           & : & 17        & : & 1             & : & 0      \\% %expt 15 --> vii
viii   &  HCN, NH$_3$, H$_2$O \& CO       & 2     & : & 1            & : & 2.5       & : & 0             & : & 2         \\% %expt 14 --> viii
\hline\\
\end{tabular}
\end{center}
$^a$ Ratios calculated from the experimental spectra.\\
\end{table}

Molecular species in the form of room temperature gas mixtures are dosed onto the gold surface by spraying at normal incidence via two injection lines. The infrared spectra of the condensed mixtures are recorded by means of Fourier Transform Reflection Absorption Infra Red Spectroscopy (FT-RAIRS) using a MCT detector in a Vertex 70 spectrometer. A typical spectrum has a 1~cm$^{-1}$ resolution and is averaged over a few tens of interferograms. 
In the gas phase, molecules in the high vacuum chamber are measured by means of mass spectroscopy using a Hiden HAL VII RGA quadrupole mass spectrometer (QMS) with a 70~eV impact electronic ionisation source.

\begin{figure*}
\centering
\includegraphics[width=\textwidth]{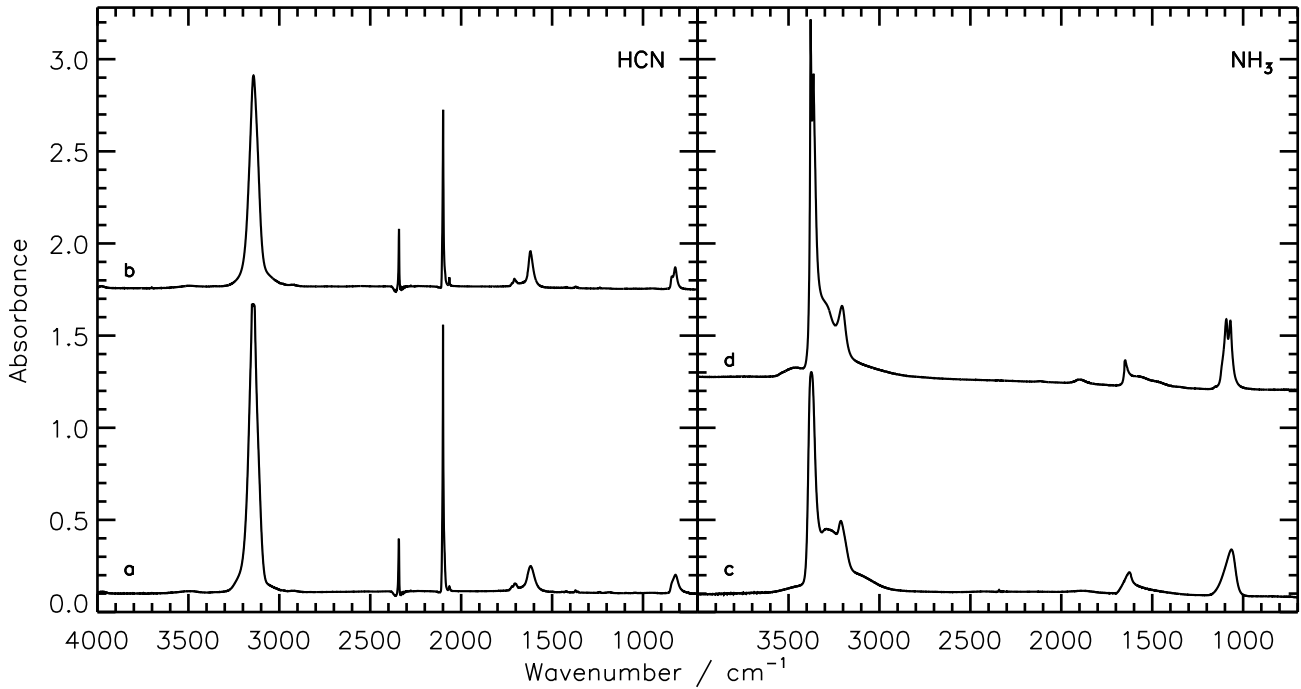}
\caption{ The infrared absorption spectra of pure HCN and NH$_3$ deposited at 10~K: a) HCN at 10~K, b) HCN after heating to 100~K, c) NH$_3$ at 10~K, d) NH$_3$ after heating to 100~K.}
\label{figure_nh3_hcn_pur}
\end{figure*}

In the experiments presented here, a series of gas-phase mixtures of HCN, NH$_3$, H$_2$O, CO and CO$_2$ were prepared in the injection lines. Ammonia is commercially available as a 99.999~$\%$ pure gas from Air Liquide. The gas phase HCN monomer was synthesised via the thermal reaction of potassium cyanide, KCN, with an excess of stearic acid, CH$_3$(CH$_2$)$_{16}$COOH, in a primary pumped vacuum line \citep{Gerakines04}. 
A small quantity of CO$_2$ was produced during HCN synthesis due to the decomposition of stearic acid.
H$_2$O vapour was obtained from deionised water purified by several freeze-pump-thaw cycles carried out under vacuum.

A complete list of experiments performed in this study is presented in Table~\ref{table1}. In all experiments, NH$_3$ was dosed via one beam line, while the other species were dosed via a second line.
Gas mixtures were dosed onto the gold surface, held at 10~K. Except where noted in the text, the surface was subsequently subjected to a heating ramp of 2~K\,min$^{-1}$, and infrared spectra were recorded at intervals of 10~K. Molecules desorbing from the surface were monitored by mass spectroscopy.

\section{Results}\label{sec-results}

\subsection{Pure HCN and NH$_3$}\label{sec-pure-hcn-nh3}

Figure~\ref{figure_nh3_hcn_pur} presents the infrared spectra of pure HCN (Experiment (i)) and pure NH$_3$ (Experiment (ii)) deposited at 10~K. The spectra have absorption bands at characteristic frequencies, which are tabulated in Table~\ref{table_freq}. The spectra of NH$_3$ \citep{dHendecourt86,Sandford93} and HCN \citep{Bernstein97,Gerakines04} are well documented in the literature; NH$_3$ is characterised by four fundamental vibrational modes $\nu_1$~--~$\nu_4$ with frequencies 3212, 1064, 3375, and 1625~cm$^{-1}$, respectively, while HCN displays vibrational modes $\nu_1$~--~$\nu_3$ with frequencies 3143, 822, and 2100~cm$^{-1}$, respectively, as well as a prominent absorption at 1617~cm$^{-1}$ assigned as the first overtone of the HCN bending mode.

The pure HCN and NH$_3$ ices were heated from 10~--~150~K at a heating rate of 2~K\,min$^{-1}$; the spectra of NH$_3$ and HCN at 100~K are also presented in Figure~\ref{figure_nh3_hcn_pur}. Between 70~--~80~K, crystallisation of NH$_3$ results in a rapid change in position and form of the four main bands: $\nu_1$ shifts to lower frequency (3206~cm$^{-1}$), $\nu_2$ and $\nu_3$ split into two peaks (1071 and 1093~cm$^{-1}$, and 3378 and 3364~cm$^{-1}$, respectively), and $\nu_4$ shifts to higher frequency (1649~cm$^{-1}$). These changes are visible in the spectrum of pure NH$_3$ at 100~K, shown in Figure~\ref{figure_nh3_hcn_pur}, trace d. The crystallisation of pure HCN occurs gradually between 10~--~120~K, with no definitive change, unlike for NH$_3$: $\nu_1$ shifts to lower frequency (3139~cm$^{-1}$) at temperatures up to 60~K, before increasing to 3144~cm$^{-1}$ by 120~K; $\nu_2$ and $\nu_3$ both shift to higher frequency above 80~K (2099 and 823~cm$^{-1}$, respectively, likely due to crystallisation of the HCN \citep{Danger11}); the overtone at 1617~cm$^{-1}$ shows similar temperature dependence as $\nu_1$, shifting to 1620~cm$^{-1}$ by 60~K then returning to 1617~cm$^{-1}$ by 120~K. Crystallisation is evinced by the difference between the low and high temperature spectra of Figure~\ref{figure_nh3_hcn_pur}.

\begin{table*}
\caption{Fundamental infrared band positions (cm$^{-1}$), for the species HCN, NH$_3$, H$_2$O, CO and CO$_2$ in various mixtures (as defined in Table~\ref{table1}). Unless otherwise stated, all values are for ices at 10~K.}
\label{table_freq}
\begin{center}
\begin{tabular}{lcccccccccc}
\hline
\multicolumn{2}{c}{Vibration mode} & Assignment & HCN      & NH$_3$    & \multicolumn{2}{c}{HCN:NH$_3$} & HCN:H$_2$O & HCN:NH$_3$:H$_2$O & \multicolumn{2}{c}{HCN:NH$_3$:H$_2$O:X}  \\
                          &                            &                     & (i)            &         (ii)       &        (iii)           &       (iv)           &       (v)             &     (vi)                    &     X = CO$_2$ (vii)$^a$     &      X = CO (viii)     \\
\hline
HCN/CN$^-$ & $\nu_3$                  & CN stretch   & 2100      & --           & 2077              &  2098            & 2091             &   2090               & 2086                                       &  2086                            \\
NH$_3$ & $\nu_3$                           & NH stretch   &  --         &  3375       &  3380             &  3398            &   --               &  3378                & 3385                                       &  3399                            \\
NH$_3$ & $\nu_2$                           & umbrella      &  --         &  1064       &  1098             &  1104            &   --               &  1101                & 1113                                       & 1112                              \\
CO$_2$ &$\nu_3$                             & CO stretch   &2343$^b$& 2342$^b$& 2342$^b$      & 2343$^b$     & 2342$^b$      & 2342$^b$          & 2340                                      & 2343$^b$                        \\
CO        &  $\nu_1$                           & CO stretch   & --           & --          & --                  & --                 & --                  & --                     & --                                          & 2138                               \\
NH$_4^+$& $\nu_4$                         & NH bend     & --           & --          & 1479              & 1471             & --                 & --                     & 1482                                       & 1498                               \\
\hline
NH$_4^+$CN$^-$~$^c$&$\nu_4$       & NH bend       & --           & --          &   1435           &  1435             & --                 &  1437             & 1437                                      &    n/a$^c$                          \\
NH$_4^+$CN$^-$~$^c$    & $\nu_1$  &  CN stretch   &   --         &  --         &    2092           &     2092         &    --              &    2092            &    2092                                  &    n/a$^c$                           \\
\hline
\end{tabular}
\end{center}
$^a$ Deposition (and spectrum measured) at 15~K. $^b$ Present as a trace pollutant (produced during the HCN synthesis). $^c$ Isolated on the surface at high temperature, after the desorption of NH$_3$ and HCN. The ice mixture containing CO (Experiment (viii)) was not heated to sufficiently high temperature to observe these bands.
\end{table*}

\begin{figure}
\centering
\includegraphics[width=0.5\textwidth]{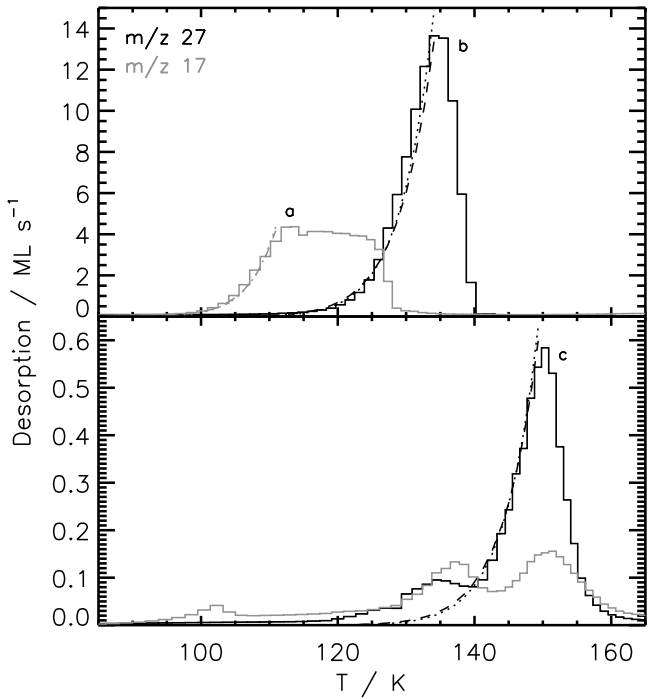}
\caption{The temperature programmed desorption spectra of a) pure NH$_3$, b) pure HCN, c) NH$_4^+$CN$^-$. The experimental desorption spectra are plotted as solid lines (m/z 27 in black and m/z 17 in grey). Overplotted on these peaks are the best fitting zeroth order desorption kinetics for each molecular species; the fixed A method is plotted as a dotted line, while the parameterised A method is plotted as a dashed line. The fitting approach is fully described in the text, where the two methods are denoted as Method 1 and Method 2, respectively.}
\label{figure_nh3_hcn_tpd}
\end{figure}

The temperature programmed desorption of the pure species was measured by quadrupole mass spectrometry. 
The major fragments (relative intensity $>$~10~\%) of NH$_3$ are the parent fragment (m/z 17, 100~\%) and the loss of one hydrogen at m/z 16 (86~\%); for HCN the major fragments are m/z 27 (100~\%) and m/z 26 (17~\%). The resulting QMS spectra of the TPD of pure NH$_3$ and pure HCN are plotted in the top panel of Figure~\ref{figure_nh3_hcn_tpd}. 

Thermal desorption is an activated process. The rate of desorption by unit surface, $r$, can be expressed by the Wigner-Polanyi equation \citep{King75}, where the desorption rate constant $k_{\text{des}}$ is described in terms of an Arrhenius law:

\begin{equation}\label{eqn:polanyi}
  r = -\frac{dN}{dt} = k_{\mbox{des}}\,N^n = A\,e^{-E_{\mbox{des}}/RT}\,N^n,
\end{equation}
where $A$ is the pre-exponential factor, $E_{\text{des}}$ is the energy of desorption of a molecule from the surface (J\,mol$^{-1}$), R is the gas constant (J\,K$^{-1}$\,mol$^{-1}$), $T$ is the temperature of the surface (K), $N$ is the number of adsorbed molecules on the surface (molecules\,cm$^{-2}$), and $n$ is the order of the reaction. The units of $A$ depend on $n$: molecules$^{1-n}$\,cm$^{-2+2n}$\,s$^{-1}$.

By rearranging Equation~(\ref{eqn:polanyi}) as:

\begin{equation}\label{eqn:polanyi2}
r = -\frac{dN}{dT} = \frac{A}{\beta}\,e^{-E_{\mbox{des}}/RT}\,N^n,
\end{equation}
where $\beta$~=~$\frac{dT}{dt}$~=~2~K\,min$^{-1}$ is the heating rate, the experimental data can be analysed. In order to calculate the energies of desorption of the isolated species HCN and NH$_3$, zeroth order desorption kinetics are assumed, as is standard practise for the multilayer desorption of bulk material \citep[see e.g.][]{Collings04,Burke10,Noble12}. In this case, A has units of molecules\,cm$^{-2}$\,s$^{-1}$. 

The energy of desorption, $E_{\text{des}}$ was calculated using two methods. 
Firstly (Method 1), assuming that the lattice vibrational frequency of the solid is 10$^{13}$~s$^{-1}$ and the number of molecules in a monolayer is approximately 10$^{15}$~cm$^{-2}$, the pre-exponential factor, $A$, can be fixed at a value of 10$^{28}$~molecules\,cm$^{-2}$\,s$^{-1}$, and the desorption energy calculated as the only free parameter in the fit of Equation~\ref{eqn:polanyi2} to the experimental data \citep{Collings03,Fuchs06}. 
Using this method, the calculated desorption energies were: $E_{\text{ads,HCN}}$~=~30~$\pm$~1~kJ\,mol$^{-1}$, and $E_{\text{ads,NH}_3}$~=~26~$\pm$~1~kJ\,mol$^{-1}$.

In the second method of fitting (Method 2), as proposed by \citet{Hasegawa92} the pre-exponential factor in Equation~(\ref{eqn:polanyi2}) was assumed to be a function of $E_{\text{ads}}$ approximated by:

\begin{equation}\label{eqn:harmonic}
  A = N_{\mbox{ML}} . \nu = N_{\mbox{ML}} \sqrt{\frac{2N_{\mbox{ML}}E_{\mbox{ads}}}{\pi^2M}},
\end{equation}
where $M$ is the mass of the adsorbate molecule, and $N_{\text{ML}}$~$\sim$~10$^{15}$~cm$^{-2}$. The advantage of this method is that the fit requires only one variable, $E_{\text{des}}$, rather than assuming or fitting the pre-exponential factor and fitting $E_{\text{des}}$ \citep{Acharyya07,Noble12}. 
Using the second method, the derived desorption energies were: $E_{\text{ads,HCN}}$~=~28~$\pm$~1~kJ\,mol$^{-1}$, and $E_{\text{ads,NH}_3}$~=~25~$\pm$~1~kJ\,mol$^{-1}$.
The pre-exponential factors were: $A_{\text{HCN}}$~=1.5~$\times$~10$^{27}$~$\pm$~5.2~$\times$~10$^{23}$~molecules\,cm$^{-2}$\,s$^{-1}$ and  $A_{\text{NH}_3}$~=~1.7~$\times$~10$^{27}$~$\pm$~2.0~$\times$~10$^{24}$~molecules\,cm$^{-2}$\,s$^{-1}$. 

For both species, the values of the desorption energies calculated by the two methods fall within the limits of uncertainty.
As discussed in \citet{Noble12b}, both methods are valid, since both pairs of parameters $\{A$, $E_{\text{des}}\}$ give the same value for the quantity of physical importance, \textit{i.e.} the $k_{\text{des}}$ desorption rate constant. We prefer to use the coupled solutions $\{A$~=~10$^{28}$~$\pm$~0~molecules\,cm$^{-2}$\,s$^{-1}$, $E_{\text{ads,HCN}}$~=~30~$\pm$~1~kJ\,mol$^{-1}\}$, and $\{A$~=~10$^{28}$~$\pm$~0~molecules\,cm$^{-2}$\,s$^{-1}$, $E_{\text{ads,NH}_3}$~=~26~$\pm$~1~kJ\,mol$^{-1}\}$.
This latter value is consistent with the published value of 25.6~$\pm$~0.2~kJ\,mol$^{-1}$ for NH$_3$ \citep{Sandford93}.

\subsection{HCN and NH$_3$ mixtures}\label{sec_hcn_nh3}

The HCN and NH$_3$ gases were prepared in separate molecular beams and co-deposited on the gold surface held at 10~K, 
with an excess of NH$_3$ in order to isolate the chemical reaction kinetics from the diffusion kinetics. 
Although this mixture (without H$_2$O) is not directly astrophysically relevant, an excess of NH$_3$ favours the complete reaction, gives HCN an homogeneous NH$_3$ environment and thus produces reaction kinetics of (pseudo-) order one in HCN.

\subsubsection{Identification of the NH$_3$~+~HCN reaction product}\label{identification}

\begin{figure*}
\centering
\includegraphics[width=\textwidth]{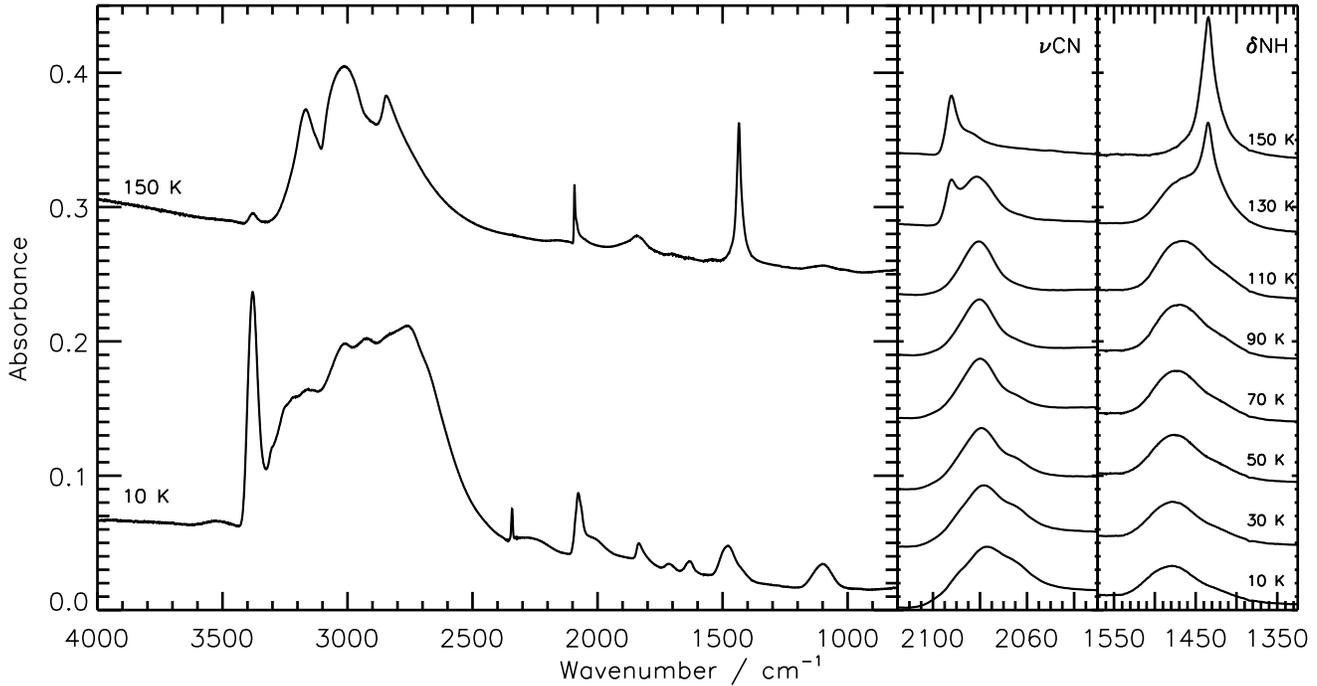}
\caption{IR spectra of Experiment (iii), a mixture of HCN and NH$_3$ with an excess of NH$_3$. Left panel: Upon deposition at 10~K, and after heating to 150~K. Middle panel: evolution of the $\nu_3$ CN stretching mode of HCN/CN$^-$ as a function of temperature. Right panel: evolution of the $\nu_4$ NH bending mode band of NH$_4^+$ as a function of temperature.}
\label{figure_nh3_hcn_melange}
\end{figure*}

The infrared spectra of Experiment (iii), a mixture of HCN and NH$_3$, are presented in Figure~\ref{figure_nh3_hcn_melange}. 
The bands attributed to HCN and NH$_3$ are slightly shifted with respect to their values in pure HCN and NH$_3$, as can be seen by comparing the values in Table~\ref{table_freq}. The presence of new bands in the spectrum at 10~K (c.f. Figure~\ref{figure_nh3_hcn_pur} and Table~\ref{table_freq}) also suggests that the species have undergone a reaction in the gas phase at 300~K during the deposition, before thermalisation to 10~K. 
It has previously been suggested that this reaction could also occur as a result of the release of energy during condensation \citep{Gerakines04}. 

The wide band at 1479~cm$^{-1}$ is attributed to the NH$_4^+$ ion, which has been extensively studied in astrophysically relevant ices \citep{Raunier03b,Raunier04,Gerakines04,GalvezApJ10}.  In addition to the broad band centred at 1479~cm$^{-1}$, which is assigned to the $\nu_4$ NH bending mode, NH$_4^+$ exhibits a broad feature in the 3000~cm$^{-1}$ region, which is the superposition of the $\nu_3$ NH stretching mode, the 2$\nu_4$ overtone, and the $\nu_2$ + $\nu_4$ combination mode \citep{Wagner50,Clutter69}. In the 3000~cm$^{-1}$ region, at low temperatures (before NH$_3$ and HCN desorption), the NH$_4^+$ contribution is difficult to deconvolve from the NH$_3$ contribution.
At 150 K, when NH$_3$ and HCN have desorbed from the surface, the IR spectrum of the product is isolated, and the NH$_4^+$ feature is revealed as a three-peaked band with maxima at 3168 ($\nu_2$ + $\nu_4$), 3014 ($\nu_3$), and 2845~cm$^{-1}$ (2$\nu_4$). 
We obtain the same  NH$_4^+$CN$^-$ IR spectrum as \citet{Gerakines04} and \citet{Clutter69}.
 
The band at 1479~cm$^{-1}$ grows and shifts redwards as the mixture of HCN and NH$_3$ is heated from 10~K to 150~K, as illustrated in the right hand panel of Figure~\ref{figure_nh3_hcn_melange}. Upon the desorption of both reactants, HCN and NH$_3$, between 110 and 150~K (as described in Figure~\ref{figure_nh3_hcn_tpd}) the band shifts to 1435~cm$^{-1}$. This shift is due to both a change in environment and to crystallisation.

The production of NH$_4^+$ necessitates the presence of an anion in the ice. 
We positively identify the counter ion as CN$^-$ from its IR band at 2092~cm$^{-1}$ (Figure~\ref{figure_nh3_hcn_melange}, middle panel). As seen for the NH$_4^+$ band at 1479~cm$^{-1}$, the HCN/CN$^-$ band at 2077~cm$^{-1}$ shifts during heating (although it shifts bluewards), and between 110 and 150~K the band narrows and shifts to 2092~cm$^{-1}$ due to a changing environment and crystallisation of the NH$_4^+$CN$^-$ product.

We identify the product of the NH$_3$~+~HCN reaction to be the NH$_4^+$CN$^-$ species with the following four arguments:
\begin{itemize} 
\item the identification of the NH$_4^+$CN$^-$ absorption bands in the IR spectrum of the HCN:NH$_3$ mixture upon heating to 150~K (Figure~\ref{figure_nh3_hcn_melange}, left hand panel), 
\item in the TPD spectrum of the HCN:NH$_3$ mixture (shown for m/z 27 in the lower panel of Figure~\ref{figure_nh3_hcn_tpd}, and discussed further in \S~\ref{desorption_rate}), part of the
  product desorbs when dominant NH$_3$ desorbs around 135~K. The mass spectrum of the pure product is obtained between 140~K and 160~K, as shown in Figure~\ref{figure_nh3_hcn_tpd}. Mass features for the product are obtained at m/z 27 (100~\%), 26 (17~\%), 17 (20~\%) and 16 (15~\%) at 150~K. The observed mass features of the product are a combination of the pure HCN and NH$_3$ desorption features. These features thus correspond to a proton transfer during the reverse reaction NH$_4^+$~+~CN$^-$~$\rightarrow$~NH$_3$~+~HCN and the subsequent rapid desorption of HCN and NH$_3$ at high temperature (140 -- 160~K). Within the observational limitations of our experiments, no m/z 44 feature corresponding to the molecular ion NH$_4^+$CN$^-$ is observed,
\item the similarity of this system to the low temperature proton transfer reaction  NH$_3$~+~HNCO~$\rightarrow$~NH$_4^+$~+~OCN$^-$  \citep{Demyk98,Raunier03a,vanBroekhuizenAA04},
\item the previous literature evidence of the spontaneous reaction between HCN and NH$_3$, represented in Equation~(\ref{eqn_hcn_nh3}), upon deposition \citep{Gerakines04,Clutter69}.
\end{itemize}
 
We can therefore assign the product of the NH$_3$~+~HCN reaction to the NH$_4^+$CN$^-$ species.

A number of values can be derived from experiments on HCN:NH$_3$ mixtures: the band strength of the CN$^-$ ion, $A_{\text{CN}}$, the activation energy of the HCN + NH$_3$ reaction, and the desorption energy of the salt NH$_4^+$CN$^-$ from the surface.

\subsubsection{The HCN bending mode at 848~cm$^{-1}$.}\label{sec_848}

\begin{figure}
\centering
\includegraphics[width=0.5\textwidth]{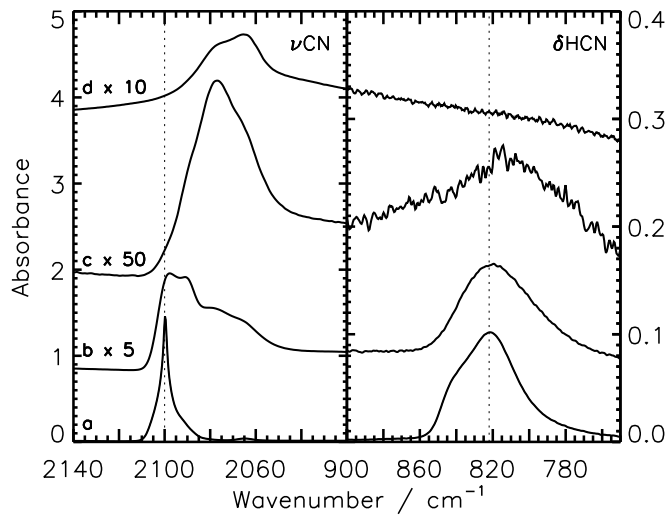}
\caption{A comparison of the CN stretching mode (left panel) and HCN bending mode (right panel) absorption bands at 10~K for a) pure HCN (Experiment~(i)), b) HCN in a low concentration of NH$_3$ (Experiment~(iv)), and c) HCN in an excess of NH$_3$ (Experiment~(iii)). Trace d is a spectrum of pure NH$_4^+$CN$^-$ at 20~K taken from \citep{Danger11}. The dotted lines mark the positions of the band maxima in Experiment~(i), and are included to guide the eye.}
\label{figure_HCNd}
\end{figure}

In a mixture of HCN:NH$_3$, the only isolated HCN band (which does not overlap with bands of NH$_3$ or CN$^-$) is the HCN bend at 848~cm$^{-1}$. Figure~\ref{figure_HCNd} (right panel) illustrates the effect of the environment on this band; spectra are plotted for Experiments~(i) pure HCN, (iv) low NH$_3$ concentration, and (iii) high NH$_3$ concentration at 10~K. The band widens and shifts redwards with increasing NH$_3$ concentration. The presence of the broad bending mode absorption feature at 800~cm$^{-1}$ in the spectrum of H$_2$O prevents the study of the HCN bending mode in mixtures containing H$_2$O. In a sample of pure NH$_4^+$CN$^-$ at 20~K (Figure~\ref{figure_HCNd}, trace d), prepared by mixing gaseous HCN and an excess of NH$_3$ in the injection lines before dosing, the HCN bending mode is not seen, as all HCN is converted to CN$^-$ before dosing onto the surface \citep{Danger11}. In all of the experiments presented here (without H$_2$O) the HCN bending mode is clearly observed, indicating that in all experiments some HCN remains unreacted on the surface at 10~K.

Conversely, the $\nu_2$ and $\nu_3$ bands of NH$_3$ both increase in frequency with the addition of higher concentrations of HCN. The $\nu_2$ band shifts from 1064~cm$^{-1}$ in pure NH$_3$ (Experiment~(ii)) to 1098~cm$^{-1}$ with a low HCN concentration (Experiment~(iii)) to 1104~cm$^{-1}$ with a higher HCN concentration (Experiment~(iv)), while the $\nu_2$ band shifts from 3375~cm$^{-1}$ to 3380~cm$^{-1}$ to 3398~cm$^{-1}$ in the same experiments. 

The absorption strength of the HCN bending mode in pure HCN at 10~K was calculated from the data in Experiment~(i) to be $A_{\delta\,\text{HCN}}$~=~1.9~$\pm$~0.7~$\times$~10$^{-18}$~cm\,molec$^{-1}$, using the area under the HCN bending mode and the HCN CN stretching mode peaks, and the absorption strength of the CN stretching mode \citep{Bernstein97}.

\subsubsection{The CN stretching mode at 2100~cm$^{-1}$.}

It is particularly interesting to note the effect of the environment on the CN stretch at 2100~cm$^{-1}$, as the band is very complex, and the contributions of the species HCN and CN$^-$ are largely irresolvable. The spectra of Experiments (i) pure HCN, (iv) low NH$_3$ concentration, and (iii) high NH$_3$ concentration at 10~K are plotted in Figure~\ref{figure_HCNd} (left panel), as for the HCN bending mode discussed in the previous section. The band at 2100~cm$^{-1}$ behaves very similarly to the HCN bending mode at 848~cm$^{-1}$, widening and shifting redwards with increasing NH$_3$ concentration. It is evident from Figure~\ref{figure_HCNd} that in the different experiments, different components contribute to the overall profile of the band. By comparison to trace d in the same figure (pure NH$_4^+$CN$^-$), it can be inferred that an increasing concentration of NH$_3$ results in an increased conversion of HCN to CN$^-$ in our mixtures at 10~K. However, due to the complexities introduced by the effects on the band position and profile of concentration and, as previously discussed with respect to Figure~\ref{figure_nh3_hcn_tpd}), temperature, it has not been possible to resolve the contributions of the species HCN and CN$^-$ to the CN stretching mode in a reliable and reproducible manner across all of our spectra.

\subsubsection{Determination of the CN$^-$ ion band strength.}

In order to determine the band strength of the CN$^-$ ion, it is necessary to calculate the CN$^-$ ion column density in our ice mixtures. Due to the difficulty of deconvolving the contributions of HCN and CN$^-$ at 2092~--~2099~cm$^{-1}$ discussed above, it is not possible to evaluate directly the number of CN$^-$ ions in the ice from its absorption band at low temperature. At temperatures above 140~K, all HCN has desorbed from the surface, and therefore the feature at 2092~cm$^{-1}$ (4.78~$\mu$m) derives solely from absorption by the CN$^-$ ion. The assumption is made that Equation~\ref{eqn_hcn_nh3} holds and therefore the number of CN$^-$ ions is equivalent to the number of NH$_4^+$ ions.  Thus, using the area under the absorption feature at 2092~cm$^{-1}$ and the number of NH$_4^+$ ions in the ice (derived from ten HCN~+~NH$_3$ experiments) the average absorption strength was calculated as $A_{\text{CN}^-, \text{140~K}}$~=~1.8~$\pm$~1.5~$\times$~10$^{-17}$~cm\,molec$^{-1}$. The uncertainty is calculated from the spread of values in the ten experiments, and is significant, reflecting the complexity of deriving such values from laboratory data. 

Using a spectrum of pure NH$_4^+$CN$^-$ at 20~K (shown in Figure~\ref{figure_HCNd}, trace d) from \citet{Danger11}, we calculated the absorption strength of CN$^-$ at 20~K as $A_{\text{CN}^-, \text{20~K}}$~=~1.8~$\pm$~0.1~$\times$~10$^{-17}$~cm\,molec$^{-1}$. This result is identical to the value calculated at high temperature, suggesting that the CN$^-$ band in pure NH$_4^+$CN$^-$ is not highly susceptible to temperature changes in the range of 20~--~140~K.

The band strength of the CN$^-$ ion is comparable to the band strength of most other interstellar molecules, unlike the OCN$^-$ ion, which exhibits a high band strength \citep[1.3~$\times$~10$^{-16}$~cm\,molec$^{-1}$][]{vanBroekhuizenAA04} that compensates for its low abundance and makes its detection possible in the IR spectra of interstellar ices. 

Further calculations were performed on data from Experiments~(iii) and (iv) in an attempt to deconvolve the contributions of HCN and CN$^-$ to the CN stretch absorption at 2100~cm$^{-1}$. Firstly, the number of HCN molecules in Experiment~(iv) was calculated from its absorption band at 848~cm$^{-1}$ (Figure~\ref{figure_HCNd}, trace b), using the absorption strength calculated in \S~\ref{sec_848} (assuming that the magnitude does not change with environment), to be $\sim$~9.3~$\times$~10$^{17}$. Secondly, using the absorption strength calculated above for $A_{\text{CN}^-, \text{20~K}}$ at 2100~cm$^{-1}$, it was possible to calculate the number of CN$^-$ radicals in Experiment~(iv), and thus infer the number of HCN molecules as $\sim$~2.5~$\times$~10$^{18}$, approximately 1.7 times higher than that calculated by the first method. The large uncertainty on this value most likely derives from the assumptions made in the calculations: that the absorption strengths of the CN$^-$ and HCN CN stretching mode absorptions, and the HCN bending mode absorption do not change with environment. As has been previously demonstrated by \citet{Borget12}, the CN stretch in nitriles is very sensitive to environment, with the absorption strength of CN in aminoacetonitrile changing by a factor of two between the pure molecule and a mixture of aminoacetonitrile and H$_2$O (1:3). Thus we conclude that, while we can accurately determine the absorption strength of CN$^-$ at low to intermediate temperatures in a pure salt, it is not possible from these data to reliably deconvolve HCN and CN$^-$ contributions to the CN stretching mode absorption at 2100~cm$^{-1}$.

The implication of this for future observations is that in the infrared spectra of ices where both HCN and CN$^-$ species are present, their bands are likely to be superimposed upon one another. Given the large difference in absorption strengths of the two species -- 5.1~$\times$~10$^{-18}$~cm\,molec$^{-1}$ for HCN \citep{Bernstein97} and 1.8~$\times$~10$^{-17}$~cm\,molec$^{-1}$ for CN$^-$ -- it is important that the band not be treated as one species, but that efforts are made to deconvolve the individual contributions of HCN and CN$^-$. Further experimental studies are required to address this complex problem.

\begin{figure}
\centering
\includegraphics[width=0.5\textwidth]{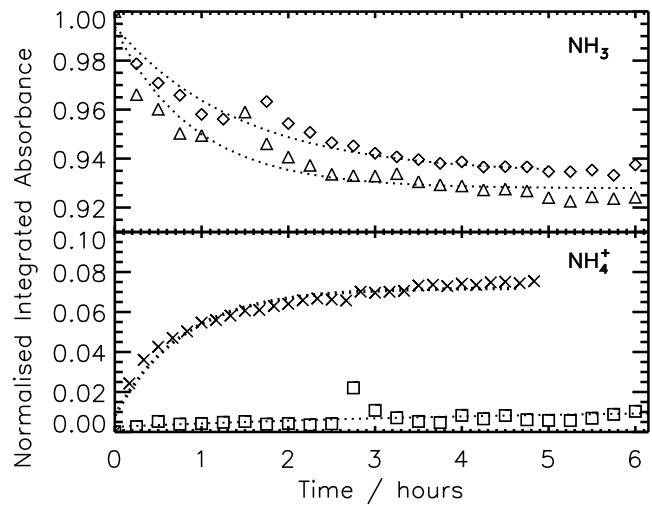}
\caption{Time evolution of the normalised abundances of the NH$_3$ reactant and the NH$_4^+$ product at different fixed temperatures. Top panel: NH$_3$ evolution at 80 K (diamond) and 90 K (triangle). Bottom panel: NH$_4^+$ evolution at 60~K (square) and 80 K (cross). Overplotted on the experimental data are the fits of Equation~\ref{kinetics}.}
\label{figure_rateA}
\end{figure}

\subsubsection{Determination of the reaction rate constant}

\begin{table}
\caption{Experimental rate constant, $k$, as a function of temperature.\label{kT}}
\begin{center}
\begin{tabular}{cc}
\hline
Temperature (K) & $k$ ($\times$~10$^{-4}$~s$^{-1}$)\\ 
\hline
60   & 0.54 $\pm$ 0.11 \\
70   & 0.91 $\pm$ 0.27 \\
80   & 2.64 $\pm$ 0.88 \\
90   & 4.35 $\pm$ 1.00 \\
100 & 4.11 $\pm$ 0.93\\
105 & 7.09 $\pm$ 0.48\\
\hline
\end{tabular}
\end{center}
\end{table}

In order to determine the activation energy of the NH$_3$~+~HCN~$\rightarrow$~NH$_4^+$CN$^-$ reaction, isothermal kinetic experiments were performed, following \citet{Bossa09}. A series of isothermal experiments was carried out in which mixtures of HCN:NH$_3$ with large excesses of NH$_3$ (the ratios were typically 1:15) were deposited at 15~K, heated as quickly as possible ($\sim$~10~K\,min$^{-1}$) to a specific temperature, and held at that fixed temperature for a number of hours, in order to study the reaction kinetics. The temperatures chosen were 60, 70, 80, 90, 100, and 105~K. Above 105~K, desorption of the reactant molecules becomes important, and below 60~K the reaction proceeds so slowly that the experiments become prohibitively long to carry out.
The abundances of NH$_3$ and NH$_4^+$ in the ice mixtures were monitored via their characteristic absorption features in the IR spectra at 3375 and $\sim$~1440~cm$^{-1}$, respectively.
The time evolution of the species' molar fractions for four experiments are plotted in Figure~\ref{figure_rateA}. The other experimental data is omitted for clarity. It is evident from this figure that an increase in temperature increases the rate of the reaction, particularly in the first few hours of each experiment.

As NH$_3$ was in excess with respect to HCN, each HCN molecule was surrounded by a homogeneous NH$_3$ environment. 
Thus, we can express the rates of destruction of NH$_3$ and formation of NH$_4^+$ with first-order reaction kinetics, in terms of the molar fraction of the reactants:

\begin{equation}
\label{kinetics} \left\{\begin{array}{rcl}
\frac{d(\mbox{HCN})}{dt} &=& -k(T).(\mbox{HCN})\\
\frac{d(\mbox{NH}_3)}{dt} &=& -k(T).(\mbox{HCN})\\
\frac{d(\mbox{NH}_4^+\mbox{CN}^-)}{dt}&=& k(T).(\mbox{HCN})\\
\end{array}\right.
\end{equation}

The first-order kinetics in HCN implies that the molar fraction of NH$_3$ is constant. Although not necessary in the first-order case, it is convenient to express the kinetic equations in terms of unitless molar fractions, which allows the calculation of a reaction rate constant $k$ in $s^{-1}$. 
Resolving Equation~\ref{kinetics} for the isothermal experimental curves gives a reaction rate constant at a fixed temperature. The fits of four of the experiments are overplotted on the experimental data in Figure~\ref{figure_rateA}.
The reaction rate constants, determined at the chosen temperatures by analysis of the NH$_4^+$ band at $\sim$~1440~cm$^{-1}$ only, are displayed in Table~\ref{kT}.

\begin{figure}
\centering
\includegraphics[width=0.5\textwidth]{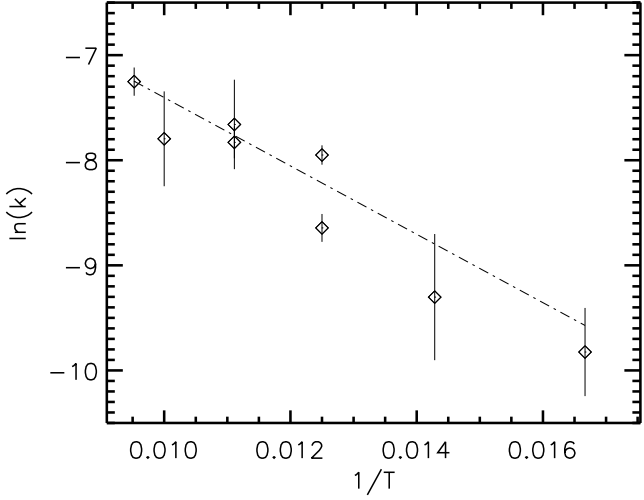}
\caption{Experimentally determined $ln(k)$ as a function of the inverse of the temperature, and the fit against an Arrhenius law (dot-dashed line). }
\label{figure_rateB}
\end{figure}

The temperature dependence of the reaction rate constant is plotted in Figure~\ref{figure_rateB}, in the form of $ln(k)$ as a function of the inverse of the temperature. Assuming $k(T)$ follows an Arrhenius law, a least-squares straight line fit to the experimental data allows the calculation of the activation energy of the reaction HCN + NH$_3$. We derive an activation energy of 2.7~$\pm$~0.4~kJ\,mol$^{-1}$ with a pre-exponential factor of 0.016$^{+0.010}_{-0.006}$~s$^{-1}$. The uncertainty on the calculated values includes the dispersion of the results measured at 80~K and 90~K, where two experiments were performed at each temperature.
The value for the activation energy is consistent with values measured previously for different thermal reaction systems \citep{Bossa09,Theule11b,Mispelaer12}.
Similarly, as observed previously, the pre-exponential factor is low, which may be due to the solvent cage effect in the solid solutions, \textit{i.e.} the need for the molecules to orientate within the solid into favourable steric configuration before reaction.
Therefore the reaction rate constant has a $k(T)$ = 0.016$^{+0.010}_{-0.006}$~s$^{-1}$\,$\exp(\frac{-2.7\pm0.4\text{~kJ\,mol}^{-1}}{RT})$ temperature dependence within the small temperature interval (60~--~105 K) where we have been able to measure it. 
The question of the extrapolation to lower temperatures is critical to interstellar ice chemistry, but difficult to address experimentally. 

\subsubsection{Determination of the NH$_4^+$CN$^-$ desorption rate constant}\label{desorption_rate}

A TPD spectrum of NH$_4^+$CN$^-$ (m/z 27 and m/z 17) is shown in the lower panel of Figure~\ref{figure_nh3_hcn_tpd}. It is evident in this TPD trace that some of the product (or unreacted HCN and NH$_3$) co-desorbs with the dominant species between $\sim$~120~--~140~K (similarly to the pure HCN in the upper panel) and that the NH$_4^+$CN$^-$ desorbs at higher temperature. 
It is not possible from the experiments presented here to determine at what temperature NH$_4^+$CN$^-$ first starts to desorb from the surface, as in none of our experiments was the NH$_4^+$CN$^-$ present on the surface with no HCN. 
Thus in all TPD traces, the signals of the desorption of HCN and NH$_4^+$CN$^-$  overlap. The dominant mass fraction for NH$_4^+$CN$^-$ is m/z 27, but as mentioned in \S~\ref{identification} above, other masses observed desorbing at this temperature include m/z 26 (17~\%), 17 (20~\%), 16 (15~\%), and 15 (2~\%). Under current experimental conditions there is no evidence of the desorption of m/z 44 or 43, and thus it is concluded that NH$_4^+$CN$^-$  desorbs from the surface entirely as HCN and NH$_3$, following the reverse reaction NH$_4^+$~+~CN$^-$~$\rightarrow$~NH$_3$~+~HCN, rather than as NH$_4^+$CN$^-$.

The zeroth-order shape of the m/z 27 NH$_4^+$CN$^-$ curve in Figure~\ref{figure_nh3_hcn_tpd} is less marked since only a few monolayers of NH$_4^+$CN$^-$ are desorbing.
Using the two methods described in \S~\ref{sec-pure-hcn-nh3} above (fitting to the m/z 27 desorption peak only, and taking the average of values calculated from five desorption spectra) the energy of desorption of NH$_4^+$CN$^-$ was calculated as $E_{\text{ads,NH}_4\text{CN}}$~=~38.0~$\pm$~1.4~kJ\,mol$^{-1}$ (Method 1) and $E_{\text{ads,NH}_4\text{CN}}$~=~35.4~$\pm$~1.3~kJ\,mol$^{-1}$ (Method 2, with $A$ = 1.3~$\times$~10$^{27}$~$\pm$~6.0~$\times$~10$^{23}$~molecules\,cm$^{-2}$\,s$^{-1}$).
When considering the energies derived for all three molecular species, the values calculated using the fixed $A$ method are consistently 6~--~7~\% higher than those calculated using the second method ($A$ as a function of $E_{\text{ads}}$).

\subsection{HCN and NH$_3$ mixtures diluted in H$_2$O}

In order to investigate the reaction of HCN and NH$_3$ under conditions relevant to ices in the interstellar medium, H$_2$O must be included in the mixtures. H$_2$O is the most abundant solid phase interstellar molecule, with abundances of 10$^{-4}$ with respect to H$_2$. 

\subsubsection{HCN:H$_2$O ice mixture}
Experiment~(v) investigates the effect of dilution of HCN in a water ice at a ratio of 1:18. It is clear from the band positions tabulated in Table~\ref{table_freq} that the peak position of the CN stretching absorption in HCN is shifted redward (with respect to that of pure HCN) when diluted in a H$_2$O ice (to 2091~cm$^{-1}$), although the shift is not as marked as that seen for the HCN/CN$^-$ band in a HCN:NH$_3$ mixture (2077~cm$^{-1}$, Experiment~(iii)). As expected, no reaction is seen for a mixture of HCN:H$_2$O, neither at 10~K nor during heating to 180~K, where both HCN and H$_2$O have desorbed.

\subsubsection{HCN:NH$_3$:H$_2$O ice mixture}

\begin{figure*}
\centering
\includegraphics[width=\textwidth]{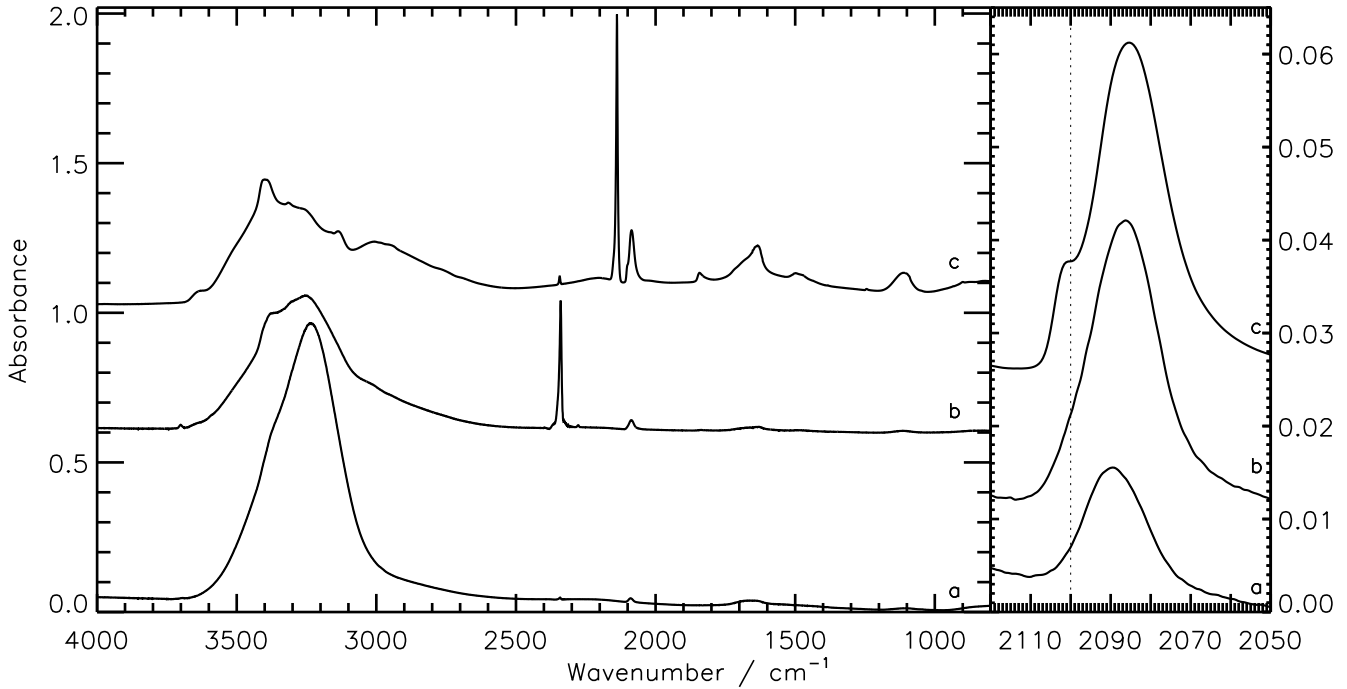}
\caption{Left panel: Infrared spectra of a) a HCN:NH$_3$:H$_2$O mixture (Experiment~(vi), 10~K), b) a HCN:NH$_3$:H$_2$O:CO$_2$ mixture (Experiment~(vii), 15~K), c) a HCN:NH$_3$:H$_2$O:CO mixture (Experiment~(viii), 10~K). Right panel: shift of the HCN/CN$^-$ band as a function of the ice mixture. The dotted line marks the position of the CN stretching mode absorption band in pure HCN.} 
\label{figure_astro_mixtures}
\end{figure*}

Experiment (vi) is a mixture of HCN:NH$_3$:H$_2$O with concentration ratio 1:3:30. This experiment investigates the HCN + NH$_3$ reaction when heavily diluted in H$_2$O, as would be expected in interstellar ices. The IR spectrum of this mixture at 10~K is shown in Figure~\ref{figure_astro_mixtures}, trace a.

The addition of H$_2$O shifts the HCN/CN$^-$ and NH$_3$ bands with respect to their positions in HCN:NH$_3$ mixtures. The HCN/CN$^-$ $\nu_3$ band is observed at 2090~cm$^{-1}$, while the NH$_3$ $\nu_2$ and $\nu_3$ bands are centred at 1101 and 3378~cm$^{-1}$, respectively.

For this mixture, the NH$_4^+$ band at 1460~cm$^{-1}$ is not observed as a distinct peak immediately after deposition at 10~K, unlike in mixtures of HCN:NH$_3$ only. There is perhaps a trace amount of NH$_4^+$CN$^-$ initially present in the mixture, but it is likely that the dilution of HCN and NH$_3$ in a H$_2$O matrix prevents their direct reaction in the gas phase during the deposition. 
During the heating of the ice mixture, a wide and shallow NH$_4^+$ band, centered at $\sim$~1490~cm$^{-1}$ develops in the spectra. At temperatures of 160~--~170~K (when almost all of the H$_2$O has desorbed from the surface), this band shifts redwards to 1437~cm$^{-1}$. 
The HCN/CN$^-$ band at 2090~cm$^{-1}$ shifts slightly redwards (6~--~7~cm$^{-1}$, to around 2086 cm$^{-1}$) during the heating of the ice mixture, due to the H$_2$O environment, the conversion of HCN to CN$^-$, the crystallisation of CN$^-$, or a combination of these effects. Upon desorption of the H$_2$O, when NH$_4^+$CN$^-$ is the only species remaining on the surface, the peak of the band shifts to 2092~cm$^{-1}$.

\begin{figure}
\centering
\includegraphics[width=0.5\textwidth]{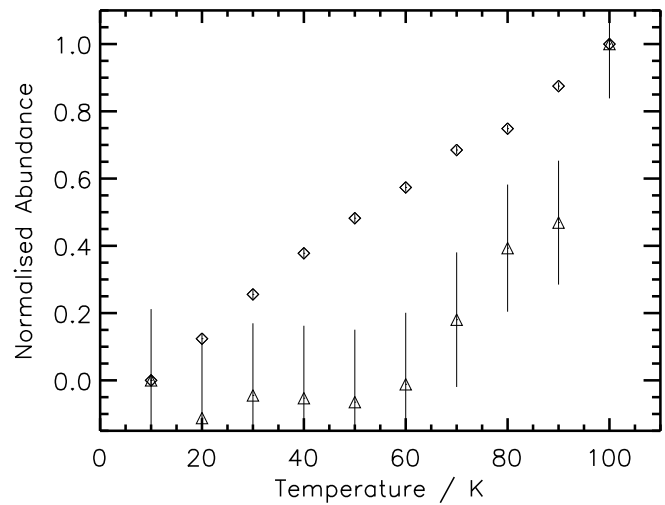}
\caption{The normalised abundance of NH$_4^+$ produced during the temperature ramps in Experiments (iii) and (vi). The reaction between HCN and NH$_3$ without H$_2$O (Experiment~(iii), diamonds) occurs more rapidly than the reaction in a H$_2$O matrix (Experiment~(vi), triangles). Values are normalised to the abundance at 100~K, before the desorption of NH$_3$ and HCN from the surface begins.}
\label{fig_h2o_matrix}
\end{figure}

\subsubsection{HCN:NH$_3$:H$_2$O:CO$_2$ and HCN:NH$_3$:H$_2$O:CO ice mixtures}

Further experiments were performed on astrophysically relevant ice mixtures (Experiments (vii) and (viii)) to investigate the effect of including CO$_2$ and CO in the ice on the position of the CN$^-$ absorption band. We wanted to investigate the possibility that the CN$^-$ ion could contribute to the band at 2041~cm$^{-1}$ (4.90 $\mu$m), assigned to OCS \citep{GibbApJS04,DartoisSSR05}, or to the so-called ``XCN'' band at 2165~cm$^{-1}$ (4.62~$\mu$m) observed towards many astrophysical environments, including high and low mass YSOs \citep[e.g.][]{Soifer79,Gibb00,vanBroekhuizen05,Aikawa12}. 
It has been previously observed that the ice absorption features of ionic species can be very sensitive to their chemical environment \citep{Schutte03} and could shift by a few wavenumbers.

The ice mixture containing CO$_2$ at 15~K is shown in Figure~\ref{figure_astro_mixtures} (left panel, trace b), while the ice mixture containing CO at 10~K is shown in trace c of the same figure. The HCN/CN$^-$ and NH$_3$ absorption bands are shifted by the presence of these two species in the ice, as tabulated in Table~\ref{table_freq}. Figure~\ref{figure_astro_mixtures} (right panel) shows the spectral region containing the HCN/CN$^-$ stretching absorption for all three astrophysically relevant ice mixtures studied. The traces are labelled as in Figure~\ref{figure_astro_mixtures} (trace a) HCN:NH$_3$:H$_2$O, trace b) HCN:NH$_3$:H$_2$O:CO$_2$, trace c) HCN:NH$_3$:H$_2$O:CO). 
The CN absorption feature of HCN/CN$^-$ is observed at 2100~cm$^{-1}$ in pure HCN, at 2091~cm$^{-1}$ when diluted into H$_2$O, and at 2086~cm$^{-1}$ in HCN:H$_2$O:CO$_2$ and HCN:H$_2$O:CO mixtures, is around 45~--~60~cm$^{-1}$ from the 2041~cm$^{-1}$ band and around 65~--~80~cm$^{-1}$ from the 2165~cm$^{-1}$ band. 

By comparison of the absorption features in Figure~\ref{figure_astro_mixtures}, and the band positions in Table~\ref{table_freq}, it is clear that the addition of H$_2$O, CO$_2$ and CO to ice mixtures containing HCN and NH$_3$ is not sufficient to induce a significant shift in the position of the HCN or CN$^-$ features. The presence of CO and CO$_2$ in the mixtures shifts the peak to slightly lower frequency (2086~cm$^{-1}$).  In the spectra presented here, it is not possible to deconvolve the contributions of HCN and CN$^-$ to the absorption feature centred at $\sim$~2090~cm$^{-1}$. No absorption features are observed in the region of 2165~--~2175~cm$^{-1}$. 

We conclude that CN$^-$ can not contribute to either the 2041~cm$^{-1}$ band, assigned to OCS, or to the ``XCN'' band.
If present in ices, it would likely be observed at or very close to 2092~cm$^{-1}$.  

Regarding the kinetics of the reaction between HCN and NH$_3$, it is evident that dilution in a H$_2$O matrix slows the formation of NH$_4^+$CN$^-$, as illustrated in Figure~\ref{fig_h2o_matrix}. Quantification of the effect of diffusion on the kinetics of a thermal reaction is extremely complicated, and will be the subject of a forthcoming article (Mispelaer et al., in preparation). In this article, we have not attempted to further study the kinetics of the HCN~+~NH$_3$ reaction in mixtures containing H$_2$O.

\section{Astrophysical implications}\label{sec-astro}

In low mass dark molecular clouds, the average abundance of gas phase HCN is around 1 -- 1.5 times higher than that of HNCO \citep[see references in][]{Vasyunina11}, and thus if OCN$^-$ is present in interstellar ices, it is reasonable to assume that HCN and CN$^-$ could equally be present.
Gas phase HCN is present in molecular clouds at abundances  (with respect to H$_2$) of $\sim$~10$^{-7}$~--~10$^{-8}$, while solid phase H$_2$O, the most abundant solid phase species, is present at abundances of $\sim$~10$^{-4}$. We can thus assume that the abundance of HCN in the solid phase is approximately 0.1~\%~H$_2$O. With values of 5.1~$\times$~10$^{-18}$~cm\,molecule$^{-1}$ \citep{Bernstein97} and 1.8~$\times$~10$^{-17}$~cm\,molecule$^{-1}$ (this work) for the CN stretch of HCN and CN$^-$, respectively, the band strengths of these molecules are not high enough to compensate for their expected low interstellar abundances. Thus, given the sensitivity of current infrared telescopes, these two molecules are unlikely to be detected in ice mantles. An HCN detection has been reported in Titan, where atmospheric chemistry has resulted in a higher abundance of solid phase HCN \citep{Burgdorf10}. Moreover, neither the HCN nor the CN$^-$ band position in different ice environments matches the 4.90~$\mu$m (2041~cm$^{-1}$) band currently assigned to OCS, nor can these species contribute to the 4.42~$\mu$m ``XCN'' band. 
However, NH$_4^+$CN$^-$ may contribute to the 6.85~$\mu$m band through its NH$_4^+$ bending absorption band, as discussed above.

\begin{figure}
\centering
\includegraphics[width=0.5\textwidth]{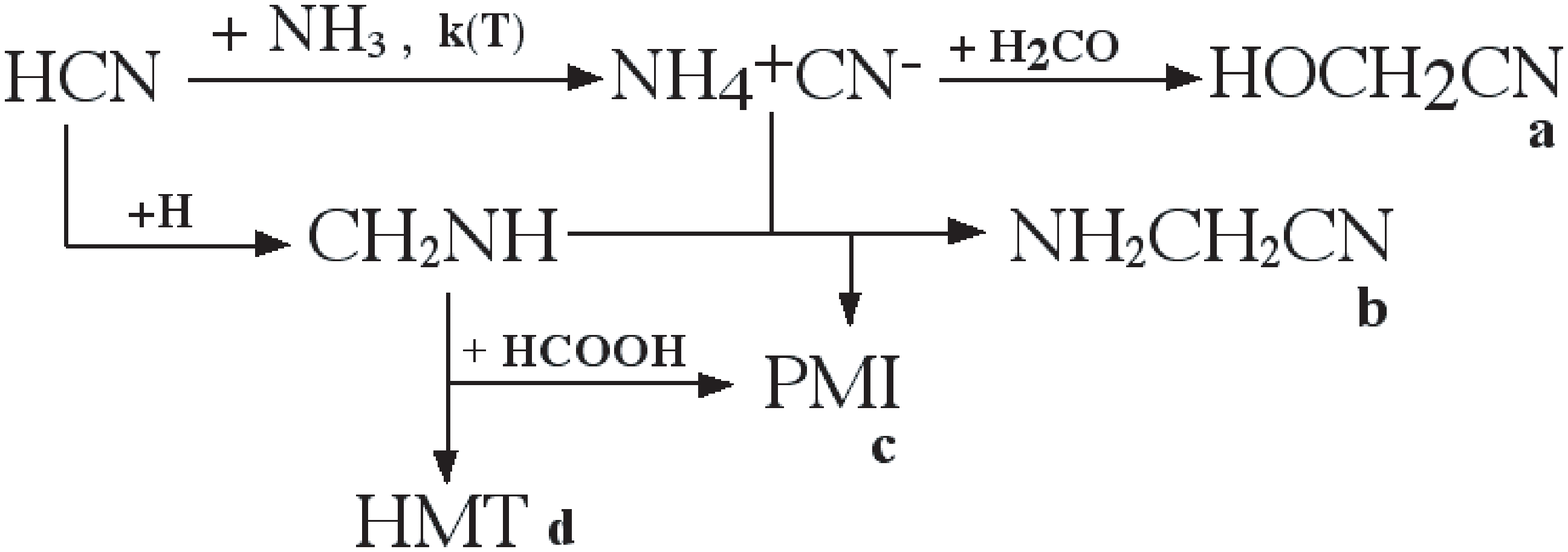}
\caption{A schematic of the simplified chemical network surrounding the NH$_3$~+~HCN~$\rightarrow$~NH$_4^+$CN$^-$ thermal reaction. The final products are a) hydroxyacetonitrile (HOCH$_2$CN), b) aminoacetonitrile (NH$_2$CH$_2$CN), c) poly(methylene-imine) (PMI, R-(CH$_2$-NH)$_n$-H, where R = HCOO or CN), and d) hexamethylenetetramine (HMT, C$_6$H$_{12}$N$_4$).}
\label{figure_hcn_network}
\end{figure}

The chemistry of HCN is of primary importance in interstellar ices, since HCN is the simplest molecular species with a CN bond and, as such, HCN is probably at the origin of some prebiotic molecules. An illustration of the chemical network centred upon HCN and NH$_4^+$CN$^-$ is shown in Figure~\ref{figure_hcn_network}. 
HCN can be produced by the hydrogenation of the CN radical, and then hydrogenated into CH$_2$NH and CH$_3$NH$_2$ \citep{Theule11a}.
As the CN$^-$ moiety of the NH$_4^+$CN$^-$ species is more nucleophilic than HCN, it could be the source of a rich ice chemistry. Its recent detection in the gas phase highlights its potential importance, and validates studies into the formation routes of interstellar ions \citep{Agundez10}.
NH$_4^+$CN$^-$ reacts with H$_2$CO to form hydroxyacetonitrile HOCH$_2$CN \citep{Danger12}.
NH$_4^+$CN$^-$  can also react with the HCN hydrogenation product CH$_2$NH to give aminoacetonitrile NH$_2$CH$_2$CN, which can go on to form glycine according to the Strecker synthesis. However, the competing reaction of NH$_4^+$CN$^-$ with CH$_2$NH can also form the poly(methylene-imine) CN-(CH$_2$-NH)$_n$-H \citep[CN-PMI, ][]{Danger11}. The reaction product depends on the starting concentrations of NH$_4^+$CN$^-$ and CH$_2$NH; an excess of CH$_2$NH promotes the polymerisation of CH$_2$NH and the production of CN-PMI, while an excess of CN$^-$ promotes the production of aminoacetonitrile (and ultimately glycine). It is likely that CN$^-$ will react with other electrophiles in interstellar ices. The reaction HCN + NH$_3$ $\rightarrow$ NH$_4^+$CN$^-$ is the preliminary step necessary to the reactions discussed above and, as such, it is important to quantify the kinetics of this rate limiting reaction.
Indeed, without H$_2$O, HCN does not react with CH$_2$O \citep{Woon01}.

It is worth emphasising that the reactions in Figure~\ref{figure_hcn_network} are thermally activated, meaning that they do not need any non-thermal process in order to occur at low temperature in astrophysical ices. 
The competition and the resulting branching ratios for these thermal reactions are entirely determined by the temperature dependence of their rate constants. As the reaction NH$_3$~+~HCN is the first step in the network, its kinetics are limiting to the overall network. 
The measured barrier to this reaction (2.7~$\pm$~0.4~kJ\,mol$^{-1}$) is a little high compared to the formation of the salt NH$_4^+$OCN$^-$ \citep[0.4~kJ\,mol$^{-1}$, ][]{Mispelaer12}, but lower than the formation of more complex species such as aminomethanol \citep[4.5~kJ\,mol$^{-1}$, ][]{Bossa09}.
This simplified network is not complete; in order to fully explain the kinetics surrounding NH$_4^+$CN$^-$, it is necessary to include all reactions, in particular those originating from NH$_3$, such as the NH$_3$~+~CO$_2$ reaction, which can be either thermally activated \citep{BossaAA08} or activated by low-energy electrons \citep{BertinPCCP09}.

In addition to constraining its reaction network, it is important to understand how much HCN can be stored in the NH$_4^+$CN$^-$ form, whether in molecular clouds, or in higher temperature environments such as comets. When the HCN hydrogenation product CH$_2$NH reacts with HCOOH, rather than with NH$_4^+$CN$^-$, the complex molecules hexamethyltetramine (HMT, C$_6$H$_{12}$N$_4$) or a second form of poly(methylene-imine) (HCOO-(CH$_2$-NH)$_n$-H) are formed. In particular, HMT is known to decompose  under VUV irradiation, producing HCN molecules and CN radicals, and therefore is a possible extended source of the observed HCN and CN in comets \citep{Bernstein95,Vinogradoff11}.

At higher temperature the reverse reaction NH$_4^+$CN$^-$~$\rightarrow$~NH$_3$~+~HCN/HNC can occur, releasing NH$_3$ and HCN (and possibly HNC) into the gas phase. This type of isomerisation has previously been shown for the HNCO/HOCN system in water ice \citep{Theule11b}. 
The formation of NH$_4^+$CN$^-$ allows HCN to be stored within a less volatile species and made available to a high temperature chemistry, when the species fragments and eventually desorbs.

As a follow up to studying its formation kinetics, it would be interesting to study the NH$_4^+$CN$^-$ photodissociation under a VUV field. 
Does this sort of irradiation release the simple molecules HCN and NH$_3$, as upon heating the salt, or rather yield a more complex chemistry? 
Moreover, the HCN/HNC ratio produced by the reverse reaction could be measured in the gas phase using gas phase spectroscopy. This type of experiment could aid interpretation of gas phase HCN/HNC ratios in interstellar environments.

The inventory of the thermal reactions involving the most abundant solid state species must be continued, and the kinetics of the rate limiting reactions should be investigated, in order to achieve a complete network of interstellar grain reactions. The competition of this series of coupled reactions must be quantified as a function of temperature and of initial abundance ratios in order to understand interstellar ice chemistry.  

\section{Conclusion}

In this work, we have shown experimentally that the purely thermal reaction between HCN and NH$_3$ produces NH$_4^+$CN$^-$ in interstellar ice analogues at low temperature. 
NH$_4^+$CN$^-$ has a characteristic CN$^-$ absorption band at 2092~cm$^{-1}$ (4.78~$\mu$m) that can be slightly redshifted by the ice environment.
However, the environmental effect is not sufficient that the CN$^-$ band can match the 2041~cm$^{-1}$ (4.90~$\mu$m) band observed in the IR spectra of interstellar ices. In addition, at low temperatures, the contributions of HCN and CN$^-$ to the observed CN stretching mode absorption are difficult to differentiate.
We have estimated the CN$^-$ ion band strength to be $A_{\text{CN}^-}$~=~1.8~$\pm$~1.5~$\times$~10$^{-17}$~cm\,molec$^{-1}$ at 20~K and at 140~K; when coupled with its expected low abundance in interstellar ices, it is unlikely that solid phase CN$^-$ could be detected by current IR telescopes. 
 We have measured the kinetics of the pure HCN and NH$_3$ reaction with an excess of NH$_3$, and we have derived the temperature dependence for the reaction rate constant to be $k(T)$ = 0.016$^{+0.010}_{-0.006}$~s$^{-1} \, \exp(\frac{-2.7\pm0.4\text{~kJ\,mol}^{-1}}{RT})$. When the reactants are diluted in water ice, the reaction is slowed down. We have also measured a desorption energy of $E_{\text{des,NH}_4\text{CN}}$~=~38.0~$\pm$~1.4~kJ\,mol$^{-1}$, assuming a pre-exponential factor of 10$^{28}$~molecules\,cm$^{-2}$\,s$^{-1}$.

\section*{Acknowledgments}

This work has been funded by the French national program Physique Chimie du Milieu Interstellaire (PCMI) and the Centre National d'Etudes Spatiales (CNES). J.A.N. is a Royal Commission for the Exhibition of 1851 Research Fellow. The authors would like to thank Prof. H. Cottin for the HCN synthesis protocol.

{}

\bsp

\label{lastpage}
\end{document}